\documentclass[10pt,letterpaper]{article}
\usepackage[top=0.85in,left=2.75in,footskip=0.75in]{geometry}

\usepackage{amsmath,amssymb}

\usepackage{changepage}

\usepackage[utf8x]{inputenc}

\usepackage{textcomp,marvosym}

\usepackage{cite}

\usepackage{nameref,hyperref}

\usepackage[right]{lineno}

\usepackage{microtype}
\DisableLigatures[f]{encoding = *, family = * }

\usepackage[table]{xcolor}

\usepackage{array}

\newcolumntype{+}{!{\vrule width 2pt}}

\newlength\savedwidth

\raggedright
\usepackage[aboveskip=1pt,labelfont=bf,labelsep=period,justification=raggedright,singlelinecheck=off]{caption}

\makeatletter
\renewcommand{\@biblabel}[1]{\quad#1.}
\makeatother

\usepackage{lastpage,fancyhdr,graphicx}
\usepackage{epstopdf}
\fancyheadoffset[L]{2.25in}
\fancyfootoffset[L]{2.25in}
\usepackage{titling,url}
\usepackage{color}
\usepackage{booktabs}
\usepackage[para,online,flushleft]{threeparttable}
\usepackage{xspace,multirow}
\usepackage{setspace,enumitem,soul}
\usepackage{framed}
\usepackage[normalem]{ulem}
\usepackage{mathtools}
\usepackage{comment,wrapfig}

\graphicspath{
{./graphs/}
{./performance/paramter_study/}
{./performance/paramter_study/additional_experiment/}
{./performance/paramter_study/all/}
{./performance/paramter_study/all_old/}
}

\newcommand{\CF}{\mbox{$\mathop{\text{\emph{CF}}}\limits$}\xspace}
\newcommand{\mCF}{\mbox{$\mathop{\text{\emph{mCF}}}\limits$}\xspace}
\newcommand{\MC}{\mbox{$\mathop{\text{\emph{MC}}}\limits$}\xspace}
\newcommand{\method}{\mbox{$\mathop{\text{\emph{DmCF}}}\limits$}\xspace}
\newcommand{\foMC}{\mbox{$\mathop{\text{\emph{foMC}}}\limits$}\xspace}
\newcommand{\ypCF}{\mbox{$\mathop{\text{\emph{ypCF}}}\limits$}\xspace}
\newcommand{\TptCF}{\mbox{$\mathop{\text{\emph{TptCF}}}\limits$}\xspace}
\newcommand{\seq}{\mbox{$\mathop{\vec{T}}\limits$}\xspace}
\newcommand{\simPY}{\mbox{$\mathop{\text{\emph{simP2Y}}}\limits$}\xspace}
\newcommand{\simYP}{\mbox{$\mathop{\text{\emph{simY2P}}}\limits$}\xspace}
\newcommand{\simP}{\mbox{$\mathop{\mathnormal{\text{sim}_{\scriptsize \patient}}}\limits$}\xspace}
\newcommand{\simY}{\mbox{$\mathop{\mathnormal{\text{sim}_{\scriptsize \physician}}}\limits$}\xspace}
\newcommand{\simT}{\mbox{$\mathop{\mathnormal{\text{sim}_{\scriptsize \term}}}\limits$}\xspace}

\newcommand{\Sdyn}{\mbox{$\mathop{\mathnormal{\text{Score}_{\text{DYN}}}}\limits$}\xspace}
\newcommand{\Scf}{\mbox{$\mathop{\mathnormal{\text{Score}_{\text{CF}}}}\limits$}\xspace}
\newcommand{\Score}{\mbox{$\mathop{\mathnormal{\text{Score}}}\limits$}\xspace}

\newcommand{\physician}{\mbox{$\mathop{\mathnormal{y}}\limits$}\xspace}
\newcommand{\Vy}{\mbox{$\mathop{\mathnormal{\mathbf{v}}}\limits$}\xspace}
\newcommand{\Set}{\mbox{$\mathop{\mathcal{S}}\limits$}\xspace}
\newcommand{\Sy}{\mbox{$\mathop{\mathcal{S}_{\scriptsize{\physician}}}\limits$}\xspace}
\newcommand{\patient}{\mbox{$\mathop{\mathnormal{p}}\limits$}\xspace}
\newcommand{\Up}{\mbox{$\mathop{\mathnormal{\mathbf{u}}}\limits$}\xspace}
\newcommand{\Sp}{\mbox{$\mathop{\mathcal{S}_{\scriptsize{\patient}}}\limits$}\xspace}
\newcommand{\term}{\mbox{$\mathop{\mathnormal{t}}\limits$}\xspace}
\newcommand{\Wt}{\mbox{$\mathop{\mathnormal{\mathbf{w}}}\limits$}\xspace}
\newcommand{\St}{\mbox{$\mathop{\mathcal{S}_{\scriptsize{\term}}}\limits$}\xspace}
\newcommand{\visit}{\mbox{$\mathop{\mathnormal{v}}\limits$}\xspace}
\newcommand{\cutoff}{\mbox{$\mathop{\text{CUTOFF}}\limits$}\xspace}

\begin{document}
\vspace*{0.2in}

\begin{flushleft}
{\Large
\textbf\newline{Improving information retrieval from electronic health records using dynamic and multi-collaborative filtering} 
}
\newline
\\
Ziwei Fan\textsuperscript{1},
Evan Burgun\textsuperscript{2},
Zhiyun Ren\textsuperscript{3},
Titus Schleyer\textsuperscript{4,5},
Xia Ning\textsuperscript{3,6*}
\\
\bigskip
\textbf{1} Department of Computer Science, University of Illinois at Chicago, Chicago, USA
\\
\textbf{2} CSCI Consulting Inc., Indianapolis, IN, USA
\\
\textbf{3} Department of Biomedical Informatics, The Ohio State University, Columbus, OH, USA
\\
\textbf{4} Regenstrief Institute, Indianapolis, IN, USA
\\
\textbf{5} Indiana University School of Medicine, Indianapolis, IN, USA
\\
\textbf{6} Department of Computer Science and Engineering, The Ohio State University, Columbus, OH, USA
\\
\bigskip

%
%





* ning.104@osu.edu

\end{flushleft}

\pagebreak

\section*{Abstract}
Due to the rapid growth of information available about
  individual patients, most physicians suffer from information overload
  when they review patient information in health information technology systems.
  In this manuscript, we present a novel hybrid dynamic and multi-collaborative
  filtering method to improve information retrieval from electronic health records.
  This method recommends relevant information from electronic health records for
  physicians during patient visits.
  It models information search dynamics using a Markov model.
  It also leverages the key idea of collaborative filtering, originating from
  Recommender Systems, to prioritize information based on various similarities
  among physicians, patients and information items.
  We tested this new method using real electronic health record data from the
  Indiana Network for Patient Care.
  Our experimental results demonstrated that for 46.7\% of testing cases,
  this new method is able to correctly prioritize relevant information
   among top-5 recommendations that physicians
  are truly interested in.





\section*{Introduction}
\label{sec:intro}

When we consider buying a book on Amazon's Website, we often benefit from items
listed in a section called ``Recommended for you.''
These recommendations, generated by a method called Collaborative Filtering
(\CF)~\cite{Ricci2015}, suggest items of possible interest based on what other
customers have viewed and purchased.
Often, these suggestions are very useful and lead to additional purchases.
However, when physicians search the electronic
health records (EHRs) with regard to a particular patient problem, the EHRs do
not make suggestions for potentially useful information.
Instead, it requires physicians to go through the same manual, cumbersome and
laborious process of searching for and retrieving information for similar
patients/problems every single time.

In this manuscript, we present \method, a novel hybrid \textbf{\emph{D}}ynamic and
\textbf{\emph{m}}ulti-\textbf{\emph{C}}ollaborative \textbf{\emph{F}}iltering method, for
information recommendation when physicians search for information from patient EHRs. 
\method integrates the following two key ideas:
\begin{itemize}[noitemsep, nolistsep,leftmargin=*]
\item collaborative filtering,
which prioritizes information items based on what similar physicians have searched for
on similar patients; and
\item dynamic modeling,
which foresees future information items of interest based on how physicians search
for information items over time.
\end{itemize}
Here, \emph{dynamics} refers to the information retrieval patterns over time (e.g., in which
order different information items are searched for; which information
item will be typically
searched for after a certain information item has been retrieved). 
\emph{Multi-collaborative filtering} (\mCF) refers to that multiple types of similarities
(e.g., physician similarities, patient similarities and information similarities)
are integrated to score information items of possible interest.  
\method models information retrieval dynamics by a first-order Markov Chain (\MC), 
and combines \MC transition probabilities (discussed in Section~\ref{sec:mc})
with \mCF scores to produce final recommendation scores for future interested information
items. 
\method recommends the information items with the highest scores to physicians. 
We tested \method on a real dataset from the Indiana Network for Patient Care (INPC).
Our experimental results demonstrate 22.3\% improvement from \method over \MC
models on top-1 recommendation (i.e., only the top recommended information item is
considered), and for 46.7\% of all the testing cases, \method is
able to correctly identify information items that are truly interested by physicians
among its top-5 recommendations.

\section*{Literature Review}
\label{sec:review}


The most relevant research to our work is from Recommender Systems, a research area that
originated in computer science. In particular, 
top-$N$ recommender systems, which recommend the top-$N$ items that are most likely to be
preferred or purchased by users,
have been used in a variety of applications in e-commerce.
%
The top-$N$ recommendation methods can be broadly classified into two
categories~\cite{Ricci2015}.
The first category is neighborhood-based collaborative filtering methods~\cite{Ning2015},
which leverage information from similar users and/or similar items to generate
recommendations. 
The second category is model-based methods, particularly
latent factor models which learn user and item latent factors and determine
user preference over items using the factors. 
%
%
%
Recent recommendation methods also include 
deep learning based approaches~\cite{Zhang2017}, in which
user preferences, item characteristics and user-item interactions
can be learned in deep architectures.
%
%

Dynamic recommender systems have been developed to recommend information of interest
over time.
Popular techniques include latent factor transition approaches~\cite{zhang2014latent},
and Markov models~\cite{sahoo2012hidden} that model 
the transitions among latent factors capturing information preference; state space
approaches~\cite{sun2012dynamic, sun2014collaborative} that model the transitions across different 
states over time; point
processes~\cite{Luo2015} 
and other statistical models ~\cite{xiong2010temporal} that learn probabilities of future events.

%
Recommendation methods have been recently used to recommend and prioritize
healthcare information,
due to the rapid growth of information available about individual patients and the
tremendous need for personalized healthcare~\cite{Pfeifer2014}.
Current applications of recommender systems in healthcare
include recommending physicians to patients on
specific diseases~\cite{Guo2016,Jiang2014}; 
recommending drugs~\cite{Wu2015},
medicine~\cite{Bao2016} and therapies~\cite{Grasser2016}; and  
recommending nursing care plans~\cite{Duan2011}, etc. 

%

\section*{Terminologies, Definitions and Notations}
\label{sec:notation}

%
\begin{wraptable}{r}{0.6\linewidth}
  \caption{Notations}
  \label{tbl:notation}
  \begin{threeparttable}
  \begin{tabular}{
      @{\hspace{5pt}}l@{\hspace{2pt}}
      @{\hspace{2pt}}p{0.7\linewidth}@{\hspace{5pt}}
    }
  \toprule
  notation & description \\ 
  \midrule
  \physician/\patient/\term/\visit	&	a physician/patient/term/visit \\
  $\seq(\physician, \patient, \visit)$ & a search term sequence of \physician on \patient in
  visit \visit \\ 

  $\Sy(\physician)$ & a set of physicians similar to \physician \\
  $\Sp(\patient)$   & a set of patients similar to \patient \\
  $\St(\term)$      & a set of terms similar to \term \\ 
  %
%


  \bottomrule
  \end{tabular}
  \end{threeparttable}

  \vspace{-10pt}
\end{wraptable}


%
In EHR systems, there is no measurement similar to numerical
rating values in Amazon that can be used to quantitatively assess how much a physician
is interested in a certain information item. 
In this case, we take a type of implicit feedback as a qualitative measurement. That is,
if a physician searches for an information item from a patient's EHR data,
the physician is considered as interested in that information item during the
diagnostic process of the patient, and that information item is useful for/relevant to the 
diagnosis of the patient. 
Thus, to evaluate whether a physician is interested in an information item on a patient,
we can check whether the physician searches for the information item from the patient's
EHR data.
Since search is typically done through submitting a search term,
we use the two terms ``search term'' and ``information item'' exchangeably, and the
problem becomes to recommend the next search term that a physician is interested in
on a certain patient. 

In this manuscript, a physician is denoted as \physician, a patient is denoted
as \patient, and a search term is denoted as \term. 
A sequence of search terms that a physician \physician searches for on
a certain patient \patient during a certain patient visit \visit is represented as
\begin{equation}
  \label{eqn:seq}
  \seq(\physician, \patient, \visit) = \{\term_{\scriptsize \visit_1} \to \term_{\scriptsize \visit_2} \to \cdots \to \term_{\scriptsize \visit_k}|\physician, \patient\},
\end{equation}
where 
$\term_{\scriptsize \visit_k}$ is the $k$-th search term during visit \visit.
Note that a physician may have multiple search sequences on a same patient during different visits.
The physician who we recommend a next search term to on a patient is referred to
as the \emph{target physician}, and the corresponding patient is referred to as the
\emph{target patient}. 
A set of physicians/patients similar to the target physician \physician/target patient
\patient is denoted as $\Sy(\physician)$/$\Sp(\patient)$, respectively. 
A set of search terms similar to a particular search term \term is denoted as $\St(\term)$.
The size of a set $S$ is denoted as $|S|$.
Additional notations will be introduced when they are used (e.g., in Section~\ref{sec:sim}).
Table~\ref{tbl:notation} presents the important notations that we use in this manuscript.

%

\section*{Overview of the Dynamic and Multi-Collaborative Filtering Method -- \method}
\label{sec:method}

%
In this manuscript, we tackle the problem of recommending the next 
search term to a physician 
while the physician is searching for information about a patient.
The key idea is to analyze search patterns in order to make
recommendations for potentially useful, other information to the physician.
To do so, we score and prioritize possible recommendations
based on the following two criteria combinatorially:
\begin{itemize}[noitemsep, nolistsep,leftmargin=*]
\item which terms the physician has searched for on the patient already and
\item which terms similar physicians have searched for on similar patients.
\end{itemize}
The first criterion considers the search dynamics under the assumption that the
past behavior of physicians is a reasonable approximation for the standard of
care~\cite{Moffett2011, Lewis2007},
and their future behavior follows a same standard of care.
Thus, future search terms can be inferred from previously searched terms and their orders. 
%
The second criterion considers patient similarities and physician similarities.
The underlying intuition is that patients share commonalities and similar patients
stimulate similar information retrieval patterns by physicians. Likewise, physicians
share commonalities which result in similar search patterns on patients. 
We propose the hybrid method \method that considers search dynamics and multiple
similarities for the next search term recommendation. 
%
%
\method consists two scoring components.
The first component is designed to address search dynamics through a
first-order Markov Chain~\cite{norris1998markov}. 
The score of a possible search term from this dynamics-based scoring component is
denoted as \Sdyn. 
The second component is to score search terms based on similarities
via multi-collaborative filtering. 
%
The score of a possible search term from this similarity-based scoring component is denoted
as \Scf.
Thus, \method
scores a next possible search term \term for a physician \physician on a patient \patient
after a sequence of searches $\seq(\physician, \patient, \visit)$ (Equation~\ref{eqn:seq})
as a linear combination of \Sdyn and \Scf, that is,
\begin{eqnarray}
  \label{eqn:score}
  \begin{aligned}
    & \Score(\term | \seq(\physician, \patient, \visit)) = (1-\alpha)\cdot\Sdyn(\term | \seq(\physician, \patient, \visit)) + \alpha\cdot\Scf(\term | \seq(\physician, \patient, \visit)),\\
  \end{aligned}
\end{eqnarray}
where $\alpha \in [0, 1]$ is a weighting parameter.
In this manuscript, if a score is generated
from a certain method $X$, a superscript $^X$ will be included on the score notation
(e.g., $\Score^{X}$, $\Score_{\text{DYN}}^{X}$ or $\Score_{\text{CF}}^{X}$). 
In general, a superscript $^X$ indicates an associated method $X$. 
All possible terms are first scored using the scoring function in
Equation~\ref{eqn:score}. The top-scored terms are recommended as the
next possible search terms. 
The first-order Markov Chain-based scoring and
the multi-collaborative filtering-based scoring
will be discussed in Section~\ref{sec:mc} and Section~\ref{sec:cf}, respectively. 
Table~\ref{tbl:method_desc} lists all the methods in the manuscript. 
%
%

\begin{table}
\begin{adjustwidth}{0in}{0in} 
  \vspace{-15pt}
  \caption{Methods}
  \label{tbl:method_desc}
  \begin{threeparttable}
  \begin{tabular}{
      @{\hspace{0pt}}l@{\hspace{2pt}}
      @{\hspace{2pt}}p{0.8\linewidth}@{\hspace{0pt}}
    }
  \toprule
  notation & method description \\ 
  \midrule

  \method	        & dynamic and multi-collaborative filtering method (Section~\ref{sec:method}) \\
  \foMC			& first-order markov chain-based scoring method (Section~\ref{sec:method:mc:scoring}) \\
  \ypCF		        & physician-patient-similarity-based \CF scoring method (Section~\ref{sec:cf:ypcf}) \\
  \TptCF		& transition-involved patient-term-similarity-based \CF scoring method (Section~\ref{sec:cf:tptcf}) \\

  \simPY		& patient-first similarity identification (Section~\ref{sec:cf:ypcf:sim:patient})\\
  \simYP		& physician-first similarity identification (Section~\ref{sec:cf:ypcf:sim:physician})\\

  \bottomrule
  \end{tabular}
  \end{threeparttable}

\end{adjustwidth}
\end{table}

\section*{Markov Chain-based Scoring}
\label{sec:mc}

\subsection*{Background on Markov Chains}
\label{sec:mc:background}

Markov Chain (\MC)~\cite{norris1998markov} represents a very fundamental dynamic modeling scheme
based on the Markovian assumption. The Markovian assumption states that in a sequence
of events
$(e_0, e_1, e_2, \cdots, e_{t-1}, e_t)$,
each event is only dependent on a small set of previous consecutive events but independent
of any earlier events. An \MC models a sequence of events so that each of the events
follows the Markovian assumption. 
The Markovian assumption is statistically represented as 
  $P(e_t|e_0, e_1, e_2, \cdots, e_{t-1}) = P(e_t|e_{t-k}, \cdots, e_{t-2}, e_{t-1})$, 
where $P(e_t|E)$ is the probability of observing event $e_t$ given the previous
event sequence $E$. 
The number of previous events that $e_t$ depends on (i.e., $k$ in $P(e_t|e_{t-k}, \cdots, e_{t-2}, e_{t-1})$)
defines the order of the \MC. A special \MC is first-order \MC, in which each event
only depends on its immediate precursor. 
%
\MC has been demonstrated to be very effective in modeling,
approximating and analyzing real-life sequence data~\cite{norris1998markov}.

%
\subsection*{First-Order Markov Chain-based Scoring -- \foMC}
\label{sec:method:mc:scoring}

We use a first-order \MC as the dynamic model to simulate the sequence of terms
that a physician \physician searches for on a patient \patient during a visit.
This method is referred to as \textbf{\emph{f}}irst-\textbf{\emph{o}}rder
\textbf{\emph{M}}arkov \textbf{\emph{C}}hain, denoted as \foMC. 
%
%
%
For a sequence
$\seq(\physician, \patient, \visit) = \{\term_{\scriptsize\visit_1}, \term_{\scriptsize\visit_2},\cdots, \term_{\scriptsize\visit_k}|\physician, \patient\}$, 
\foMC calculates a dynamics-based score $\Score_{\text{DYN}}^{\scriptsize\foMC}$
of a next possible search term \term after $\term_{\scriptsize\visit_k}$
as the transition probability from 
$\term_{\scriptsize\visit_k}$ to \term, that is,
\begin{equation}
  \label{eqn:mc_sdyn}
  \Score_{\text{DYN}}^{\scriptsize\foMC}(\term|\seq(\physician, \patient, \visit) ) = P(\term|\term_{\scriptsize \visit_k}), 
\end{equation}
where $P(\term|\term_{\scriptsize \visit_k})$ is the transition probability from
$\term_{\scriptsize \visit_k}$ to $\term$ in a first-order \MC. 
The transition probability  $P(\term_j|\term_i)$ from a term $\term_i$ to another term
$\term_j$ in a first-order \MC is calculated as the ratio of the total 
frequency of transitions from $\term_i$ to $\term_j$ over the total frequency of all
transitions from $\term_i$ to any terms, that is,
%
%
\begin{equation}
  \label{eqn:mc_p}
  P(\term_j|\term_i) =
  \bigg[\sum\limits_{\scriptsize{\seq(\physician, \patient, \visit)}}h(\term_i \to \term_j| \seq(\physician, \patient, \visit))\bigg]
  \Bigg/
  \bigg[\sum\limits_{\scriptsize{\seq(\physician, \patient, \visit)}}~\sum\limits_{\scriptsize{(\term_i\to\term_k) \in \seq(\physician, \patient, \visit)}}h(\term_i \to \term_k|\seq(\physician, \patient, \visit))\bigg], 
\end{equation}
where $(\term_i\to\term_k) \in \seq(\physician, \patient, \visit)$ represents that
$(\term_i \to \term_k)$ is in $\seq(\physician, \patient, \visit)$,
$h(\term_i \to \term_j| \seq(\physician, \patient, \visit))$
is the frequency of the transitions from $\term_i$ to $\term_j$ in
$\seq(\physician, \patient, \visit)$. 
%
Thus, $\Score_{\text{DYN}}^{\scriptsize\foMC}$ as in
Equation~\ref{eqn:mc_sdyn} is not specific to a particular
physician or patient, but corresponds to 
clinical practices that are summarized from
all available physicians and patients. 


\section*{Multi-Collaborative Filtering-based Scoring}
\label{sec:cf}

\subsection*{Background on Collaborative Filtering}
\label{sec:cf:background}

Collaborative Filtering (\CF) is a popular technique in Recommender Systems~\cite{Ricci2015}
for recommending items to a target user.
The fundamental idea of \CF is that ``similar users like similar items''. 
User-based \CF methods first identify similar users to the target user,
and then recommend to the target user the items that are preferred by similar users.
Item-based \CF methods first identify 
items similar to the
target user's preferred items, and then recommend to the target user such similar items. 
Thus, \CF methods heavily depend on the calculation of user similarity and item similarity. 
A typical way to calculate user similarity is to represent each user using her
preference profile over items, and calculate user similarity as the item preference profile
similarity.
Likewise, a typical way to calculate item similarity is to represent each item using its
preference profiles across users, and calculate item similarity as the user preference
profile similarity.
The user similarity function and item similarity function in \CF are often pre-defined,
and thus the recommendations based on similarities can be easily interpreted. 
\CF is particularly powerful when user and item data are sparse, which is often the case in
real-life applications. 
%
%
\CF is also well-known for its scalability on large-scale problems, particularly when the
user
similarity and item similarity can be calculated in parallel trivially.

\subsection*{Physician-Patient-Similarity-based CF Scoring -- \ypCF}
\label{sec:cf:ypcf}

We developed a \CF method that generates search term
recommendations from similar physicians and patients.
This method first identifies similar physicians and similar patients
(discussed in Section~\ref{sec:cf:ypcf:sim}) and then scores
terms searched by similar physicians on similar patients
(discussed in Section~\ref{sec:cf:ypcf:cf}). 
This method is referred to as ph\textbf{\emph{y}}sician-\textbf{\emph{p}}atient-similarity-based
\textbf{\emph{C}}ollaborative \textbf{\emph{F}}iltering, and denoted as \ypCF.
%

\subsubsection*{Identifying similar physicians and similar patients}
\label{sec:cf:ypcf:sim}

We developed two approaches to identifying the set of similar physicians and
the set of similar patients, depending on which set is identified first. 

\textbf{Patient-First Similarity Identification -- \simPY}
\label{sec:cf:ypcf:sim:patient}
%
%
In the first approach, a set of patients similar to the target patient \patient is first
identified, and then based on the similar patients, a set of physicians similar to the target
physician \physician is then selected. 
This approach is denoted as \simPY (i.e., from \textbf{\emph{P}}atients to ph\textbf{\emph{Y}}sicians).
%
In \simPY, the set of patients similar to the target patient \patient is represented as
\begin{equation}
  \label{eqn:multicf:sim:patient}
  \Set_{\scriptsize{\patient}}^{\text{P2Y}}(\patient) = \{\patient_1, \cdots, \patient_{k_p}|\patient\}, 
\end{equation}
and is composed of
the \mbox{top-$k_p$} most similar patients to the target
patient \patient (patient-patient similarity will be
discussed later in Section~\ref{sec:sim}). 
Given $\Set_{\scriptsize \patient}^{\text{P2Y}}(\patient)$, a set of physicians similar
to the target physician \physician is represented as
\begin{equation}
  \label{eqn:multicf:sim:patient:physician}
  \Set_{\scriptsize{\physician}}^{\text{P2Y}}(\physician|\patient) = \{\physician_1, \cdots, \physician_{k_y}|\Set_{\scriptsize \patient}^{\text{P2Y}}(\patient)\}, 
\end{equation}
and selected as follows:
first, physicians who have ever searched for same terms
on \patient and on one or more patients in $\Set_{\scriptsize \patient}^{\text{P2Y}}(\patient)$ are identified.
From such physicians, the \mbox{top-$k_y$} most similar physicians to \physician 
are selected into $\Set_{\scriptsize \physician}^{\text{P2Y}}(\physician|\patient)$
(physician-physician similarity
will be discussed later in Section~\ref{sec:sim}). 
%

\textbf{Physician-First Similarity Identification -- \simYP}
\label{sec:cf:ypcf:sim:physician}
%
The second approach is to first identify a set of physicians similar to the
target physician \physician, and then based on the similar physicians, to identify a
set of similar patients. This approach is denoted as \simYP (i.e., from
ph\textbf{\emph{Y}}sicians to \textbf{\emph{P}}atients). 
In \simYP, the set of similar physicians is represented as 
\begin{equation}
  \label{eqn:multicf:sim:physician}
  \Set_{\scriptsize \physician}^{\text{Y2P}}(\physician) = \{\physician_1, \cdots, \physician_{k_y}| \physician\},
\end{equation}
and has the \mbox{top-$k_y$} most similar physicians to
\physician.
Based on $\Set_{\scriptsize \physician}^{\text{Y2P}}(\physician)$, 
a set of patients similar to the target patient \patient, denoted as
\begin{equation}
  \label{eqn:multicf:sim:physician:patient}
  \Set_{\scriptsize \patient}^{\text{Y2P}}(\patient|\physician) = \{\patient_1, \cdots, \patient_{k_p}|\Set_{\scriptsize \physician}^{\text{Y2P}}(\physician)\},
\end{equation}
is identified as patient \patient's \mbox{top-$k_p$} most similar patients
on whom physicians in $\Set_{\scriptsize \physician}^{\text{Y2P}}(\physician)$
have ever searched for same terms as on \patient. 
%
%
%

\subsubsection*{Collaborative Filtering in \ypCF}
\label{sec:cf:ypcf:cf}

From $\Sy(\physician)$ and $\Sp(\patient)$
(either $\Set_{\scriptsize \patient}^{\text{P2Y}}(\patient)$
and $\Set_{\scriptsize \physician}^{\text{P2Y}}(\physician|\patient)$, 
or
$\Set_{\scriptsize \physician}^{\text{Y2P}}(\physician)$
and
$\Set_{\scriptsize \patient}^{\text{Y2P}}(\patient|\physician)$), 
a set of physician-patient-term triplets, denoted as
$\Set_{\scriptsize{\physician\patient\term}}^{\scriptsize \ypCF}(\Sy(\physician), \Sp(\patient)) = \big\{\langle\physician_i, \patient_j, \term_k\rangle|\physician_i \in \Sy(\physician), \patient_j \in \Sp(\patient), \term_k \in \seq(\physician_i, \patient_j, \visit_l), \forall \visit_l\big\}$, is constructed. 
%
%
That is, $\Set_{\scriptsize{\physician\patient\term}}^{\scriptsize \ypCF}(\Sy(\physician), \Sp(\patient))$ has all the
$\langle\physician_i, \patient_j, \term_k\rangle$ 
triplets such that physician $\physician_i\in\Sy(\physician)$ has searched for term
$\term_k$ for patient $\patient_j \in \Sp(\patient)$. 
Thus, for a sequence
$\seq(\physician, \patient, \visit) = \{\term_{\scriptsize\visit_1}, \term_{\scriptsize\visit_2},\cdots, \term_{\scriptsize\visit_k}|\physician, \patient\}$, 
the score $\Score_{\text{CF}}^{\scriptsize\ypCF}$ of
a next possible search term \term is calculated as follows:
%
%
%
\begin{equation}
  \vspace{-5pt}
  \label{eqn:ypscf}
\begin{aligned}
 \Score_{\text{CF}}^{\scriptsize\ypCF} & (\term|\seq(\physician, \patient, \visit))  = \bar{f}(\langle\physician, \patient, \cdot\rangle) + \\
 &    {\sum\limits_{\mathclap{\scriptsize\langle\physician', \patient', \term\rangle \in \Set_{\scriptsize{\physician\patient\term}}^{\scriptsize \ypCF}}} \hat{f}(\physician', \patient', \term)
      \cdot\simY(\physician, \physician')\cdot\simP(\patient, \patient')}
    \Bigg/
    {\sum\limits_{\mathclap{\scriptsize{\substack{\physician', \patient': \\\exists \langle\physician', \patient', \term\rangle \in \Set_{\scriptsize{\physician\patient\term}}^{\scriptsize \ypCF}}}}}\simY(\physician, \physician')\cdot\simP(\patient, \patient')}, 
\end{aligned}
\end{equation}
where 
%
\begin{doublespace}
$
  \bar{f}(\langle\physician, \patient, \cdot\rangle) =
      {\sum_{{\scriptsize \term: \langle\physician, \patient, \term \rangle \in \Set_{\scriptsize{\physician\patient\term}}^{\scriptsize \ypCF}}}f({\small \langle\physician, \patient, \term\rangle})}
      \big/~~
      {\sum_{{\scriptsize \term: \langle\physician, \patient, \term \rangle \in \Set_{\scriptsize{\physician\patient\term}}^{\scriptsize \ypCF}}} 1},
      $
%
and
\end{doublespace}
%
\noindent
$
\hat{f}(\langle\physician', \patient', \term\rangle) = f(\langle\physician', \patient', \term\rangle) - \bar{f}(\langle\physician', \patient', \cdot\rangle),
$
%
$f(\langle\physician', \patient', \term \rangle)$ is the frequency of
the triplet $\langle\physician', \patient', \term \rangle$ (i.e., how many times
$\physician'$ searches for $\term$ on $\patient'$ in total);
$\bar{f}(\langle\physician, \patient, \cdot\rangle)$ 
is the average frequency of all possible terms that $\physician$ searches for on $\patient$;
$\hat{f}(\langle\physician, \patient, \cdot\rangle)$ 
is the centered frequency for $\langle\physician, \patient, \cdot\rangle$ (i.e., shifted
by $\bar{f}(\langle\physician, \patient, \cdot\rangle)$)
in order to reduce the bias from searches with different frequencies; 
and $\simY(\physician, \physician')$ and $\simP(\patient, \patient')$ are the similarity
between $\physician$ and $\physician'$, and the similarity between $\patient$ and $\patient'$,
respectively (discussed in Section~\ref{sec:sim}). 
The intuition behind the scoring scheme in Equation~\ref{eqn:ypscf} is that
the possibility that \physician searches for \term on \patient after a sequence of
searches is the aggregation of
1). the average possibility of \physician searching for
arbitrary search terms (i.e., the first term in Equation~\ref{eqn:ypscf}), and
2). the possibility that similar physicians search for \term on similar
patients (i.e., the second term in Equation~\ref{eqn:ypscf}).
%

%

\subsection*{Transition-Involved Patient-Term-Similarity-based CF Scoring -- \TptCF}
\label{sec:cf:tptcf}

The order in which a physician searches for different terms
could indicate a diagnosis process, and therefore the search order deserves
additional consideration.
We developed a new patient-term-similarity-based \CF scoring method that 
involves the transitions among search terms.
%
Patient similarities and term similarities are considered in this method,
which is different from those in \ypCF
(i.e., physician similarities and patient similarities in \ypCF). 
This method is referred to as \textbf{\emph{T}}ransition-involved
\textbf{\emph{p}}atient-\textbf{\emph{t}}erm-similarity-based \textbf{\emph{C}}ollaborative \textbf{\emph{F}}iltering, denoted
as \TptCF. 
\TptCF aggregates  from all similar patients
the transitions from the last search term in a sequence $\seq(\physician, \patient, \visit)$
(Equation~\ref{eqn:seq}) to another search term.
Specifically, \TptCF identifies a set of patients $\Sp(\patient)$ similar to the
target patient \patient and a set of terms $\St(\term_{\scriptsize{\visit_k}})$
similar to the last search term $\term_{\scriptsize \visit_k}$ in
$\seq(\physician, \patient, \visit)$.
The set $\St(\term_{\scriptsize{\visit_k}})$ contains the terms with term-term
similarity (discussed in Section~\ref{sec:sim}) to $\term_{\scriptsize \visit_k}$
above a threshold $\beta$. 
Then \TptCF looks into what physicians search for on patients in $\Sp(\patient)$ after
they searched for a similar term in $\St(\term_{\scriptsize{\visit_k}})$.
The underlying assumption is that similar patients stimulate similar patterns of
search sequences. 
Thus, the score $\Score_{\text{CF}}^{\scriptsize \TptCF}$ of a next possible
search term \term is calculated as follows:
\begin{eqnarray}
  \label{eqn:tdscf}
  \displaystyle{
  \begin{aligned}
   \Score_{\text{CF}}^{\scriptsize \TptCF} & (\term|\seq(\physician, \patient, \visit)) = \\
    & \sum\limits_{\scriptsize \patient'\in \Sp(\patient)}\big\{\frac{\simP(\patient, \patient')}{\sum\limits_{{\scriptsize \patient'' \in \Sp(\patient)}} \simP(\patient, \patient'')} 
    \times  \sum\limits_{\scriptsize {\term' \in \St(\term_{\tiny{\visit_k}})}}\frac{g(\term'\to\term|\patient')\simT(\term_{\scriptsize \visit_k}, \term')}{\sum\limits_{{\scriptsize \term'' \in \St(\term_{\tiny{\visit_k}})}} g(\term''\to\term|\patient')}\big \},
      \\
  \end{aligned}
  }
\end{eqnarray}
where 
$g(\term'\to\term|\patient')$ is the frequency of transitions from term $\term'$ to
term \term for patient $\patient'$ from all possible searches on $\patient'$,
$\simT(\term_{\scriptsize \visit_k}, \term')$ is the term-term similarity between
$\term_{\scriptsize \visit_k}$ and $\term'$ (discussed in Section~\ref{sec:sim}).


\section*{Similarity Calculation}
\label{sec:sim}

\textbf{Physician-Physician Similarities -- \simY}
%
We first represent each physician \physician using a vector of search term frequencies,
denoted as \Vy. Each dimension of \Vy corresponds to a term, and the value in each
dimension of \Vy is the total frequency that the corresponding term has been searched by
\physician.
Note that the frequency is aggregated from all the patients that \physician
searches on. This representation scheme is very similar to the bag-of-word
representation in text mining~\cite{Aggarwal2012}. 
Given the representation, the similarity between two physicians \physician and
$\physician'$ is calculated as
the cosine similarity between $\Vy_{\scriptsize \physician}$
and $\Vy_{\scriptsize \physician'}$, that is, 
\begin{equation}
  \label{eqn:simy}
  \simY(\physician, \physician') = \cos(\Vy_{\scriptsize \physician}, \Vy_{\scriptsize \physician'}). 
\end{equation}
The intuition is that the search term distribution
indicates physician specialties and expertise, and physicians of similar specialties and
expertise are considered similar.

\textbf{Patient-Patient Similarities -- \simP}
%
Similarly as for physicians, each patient is also
represented using a vector of term frequencies,
denoted as \Up. Each dimension of \Up corresponds to a term, and the value in each
dimension of \Up is the total frequency of the corresponding term searched for by all
physicians. The term distribution represents the health histories of the patient, and
thus a reasonable patient representation. Given the representation, the similarity
between two patients \patient and $\patient'$ is calculated as the cosine similarity
between $\Up_{\scriptsize \patient}$ and $\Up_{\scriptsize \patient'}$, that is,
\begin{equation}
  \label{eqn:simp}
  \simP(\patient, \patient') = \cos(\Up_{\scriptsize \patient}, \Up_{\scriptsize \patient'}). 
\end{equation}
%

\textbf{Term-Term Similarities -- \simT}
%
Each term \term is represented using a vector of patient frequencies, denoted as
\Wt. Each dimension in \Wt corresponds to a patient, and the value in each dimension
of \Wt is the total frequency that term \term is searched for by all physicians. 
The term-term similarity between terms \term and $\term'$ is calculated as the cosine
similarity between $\Wt_{\scriptsize \term}$ and $\Wt_{\scriptsize \term'}$, that is,
\begin{equation}
  \label{eqn:simt}
  \simT(\term, \term') = \cos(\Wt_{\scriptsize \term}, \Wt_{\scriptsize \term'}). 
\end{equation}
The underlying assumption is that if two terms are frequently searched for on a same patient,
they are considered as similar in their medical meanings and relatedness.

\section*{Materials}
\label{sec:materials}

\subsection*{Data}
\label{sec:materials:data}

%

\begin{table}[!ht]
\begin{adjustwidth}{0in}{0in} 
  \caption{Statistics of INPC Dataset}
  \label{tbl:dataset}
  \begin{threeparttable}
    \begin{tabular}{
	@{\hspace{5pt}}l@{\hspace{5pt}}
        @{\hspace{5pt}}r@{\hspace{5pt}}
        @{\hspace{5pt}}r@{\hspace{5pt}}
        @{\hspace{5pt}}r@{\hspace{5pt}}
        @{\hspace{5pt}}r@{\hspace{5pt}}
        @{\hspace{5pt}}r@{\hspace{0pt}}
        @{\hspace{5pt}}r@{\hspace{0pt}}
        @{\hspace{5pt}}r@{\hspace{5pt}}
      }
      \toprule
      dataset   &  \#\patient  & \#\physician & \#\term & \#\seq &
      len(\seq) & len(\seq)/\#\patient  & len(\seq)/\#\seq \\
      \midrule
      INCP               & 13,819 & 2,121 & 9,781 & 24,183 & 69,770 & 5.049 & 2.885 \\
      \cutoff (training) &  8,471 & 1,542 & 6,550 & 13,677 & 38,553 & 4.551 & 2.819 \\ 
      \cutoff (testing)  &   624  & 147   &   654 &    692 &  2,506 & 4.016 & 3.621 \\
      \bottomrule
    \end{tabular}
    \begin{tablenotes}
      \setlength\labelsep{0pt}
      \begin{footnotesize}
      \item
	In this table, \#\patient is the number of patients;
        \#\physician is the number of physicians;
        \#\term is the number of terms;
        \#\seq is the number of sequences;
        len(\seq) is total length of sequences;
        len(\seq)/\#\patient is average length of sequences per patient
        and len(\seq)/\#\seq is average length of sequences. 
        \par
      \end{footnotesize}
    \end{tablenotes}
  \end{threeparttable}

 \vspace{-10pt}
\end{adjustwidth}
\end{table}

The data we use for experiments come from the Indiana Network for Patient Care (INPC)
\footnote{IRB Protocol \# 1612682149 ``Supporting information retrieval in the ED through collaborative filtering''.}.
The INPC is Indiana's major health information exchange, and offers physicians
access to the most complete, cross-facility virtual electronic patient records in the nation.
Implemented in the 1990s, the INPC collects data from over 140 Indiana hospitals,
laboratories, long-term care facilities and imaging centers.
We extracted the INPC search logs that were generated between 01/24/2013 to 09/24/2013. 
Table~\ref{tbl:dataset} presents the statistics of the INPC dataset.
Figure~\ref{fig:seq_len_dist} presents the distribution of sequence length in the
dataset.
It is notable that search sequences are typically very short (on average 2.89 search terms
per each sequence).
Figure~\ref{fig:term_dist} presents the distribution of the number of
unique terms for each patient. On average, each patient has
3.85 unique search terms.
The short sequences and small number of unique search terms per patient make the
recommendation problem difficult, because the available data 
are very sparse. 

\begin{figure}[!h]
  \centering
  \begin{minipage}{.48\textwidth}
    \centering
    {\input{graphs/seq_length_hist.tex}}
    \captionof{figure}{Distribution of INPC sequence length}
    \label{fig:seq_len_dist}
  \end{minipage}%
  \hfill
  \begin{minipage}{.48\textwidth}
    \centering
    {\input{graphs/numuniqitems_pat_hist.tex}}
    \captionof{figure}{Distribution of INPC \# unique terms per patient}
    \label{fig:term_dist}
  \end{minipage}
\end{figure}

%

\subsection*{Experimental Protocols and Evaluation Metric}
\label{sec:materials:protocol}

%
We use the following experimental
protocol to evaluate our methods on the INPC dataset:
all the search sequences are split by the same
cut-off time. Any searches before the cut-off time are in the training set, and
any searches after the cut-off time are in the testing set.
The models are trained using only training set, for example, the transition probabilities
(Equation~\ref{eqn:mc_p}) are constructed only using the search sequences and terms in
training set, and the various similarities (Equation~\ref{eqn:simy}, \ref{eqn:simp}
and \ref{eqn:simt}) are calculated only from the training set. 
This protocol is referred to as cut-off cross validation, denoted as \cutoff. 
Figure~\ref{fig:cutoff_graph} demonstrates the \cutoff experimental protocol.
%

%
\begin{figure}[!h]
  \centering
  \vspace{-10pt}
  \begin{framed}
    \vspace{-25pt}
           {\input{graphs/cutoff_graph.tex}}
           \vspace{-5pt}
           \captionof{figure}{\cutoff experimental protocol}
           \label{fig:cutoff_graph}
           \vspace{2pt}
  \end{framed}
  \vspace{-15pt}
\end{figure}
%

We use the cut-off time 08/15/2013
(additional results for other cut-off times are available
in the supplementary materials\footnote{\url{https://cs.iupui.edu/~zifan/sub.pdf}}).
%
This cut-off time is selected because 
sufficient search terms from a majority of the search sequences are retained in training set
before the cut-off time and
meanwhile sufficient search sequences have testing terms after the cut-off time.
After the split, the statistics for the training and testing data
is presented in Table~\ref{tbl:dataset} (in ``\cutoff'' rows).
%
This \cutoff setting is close to the realistic scenario,
that is, all the data before a certain time should be used to predict information after that
time.
However, a shortcoming of \cutoff is that many early search sequences may not have testing
terms, and many late search sequences will not have anything in the training set.
Sequences that do not have testing terms are still used to train models.
Sequences that do not have training terms are not used.
For those sequences which have terms after the cut-off time, only the first one of
the terms after the cut-off time will be used
for evaluation. 

The model performance is measured using Hit-Rate at $N$ (HR@$N$). 
For a sequence, a hit is defined as a recommended term that is truly the next search
term. 
HR@$N$ is the percentage of testing sequences that have a hit and the hit appears
among the top-$N$ recommended terms. 
Higher HR@$N$ values indicate better performance.

\section*{Experimental Results and Discussions}
\label{sec:experiments}

\subsection*{Overall Performance}
\label{sec:experiments:overall}
\begin{table}[!ht]
\begin{adjustwidth}{-0.8in}{0in} 
  \centering
  \caption{\mbox{Overall Performance Comparison with \cutoff (08/15/2013)}}
  \label{tbl:results:cuttof:0815}
  \begin{small}
  \begin{threeparttable}
      \begin{tabular}{
	@{\hspace{7pt}}c@{\hspace{7pt}}
        @{\hspace{7pt}}c@{\hspace{7pt}}
        @{\hspace{7pt}}r@{\hspace{7pt}}
        @{\hspace{7pt}}r@{\hspace{7pt}}
        @{\hspace{7pt}}r@{\hspace{7pt}}
        @{\hspace{7pt}}r@{\hspace{7pt}}
        @{\hspace{0pt}}r@{\hspace{0pt}}
        @{\hspace{7pt}}r@{\hspace{7pt}}
        @{\hspace{7pt}}r@{\hspace{7pt}}
        @{\hspace{7pt}}r@{\hspace{7pt}}
        @{\hspace{7pt}}r@{\hspace{7pt}}
        @{\hspace{7pt}}r@{\hspace{7pt}}
        }        
        \toprule
        method   & sim & $\alpha$ & $|\Sp|$ & $|\Sy|$ & $\beta$ & & HR@1 & HR@2 & HR@3 & HR@4 & HR@5 \\
        \midrule
        \foMC  & -                       & -          & - & - & - & & \textbf{0.202}         & \textbf{0.297}           & \textbf{0.338}           & \textbf{0.378}   & \textbf{0.393}\\

        \midrule
        \multirow{8}{*}{\ypCF}  & \multirow{3}{*}{\simPY} & - & $   1$ & $   1$  & - & & \textbf{0.249} & \textbf{0.355} &   \textbf{0.406}  &  0.417 &  0.428 \\
                                &                         & - & $  50$ & $   2$  & - & & 0.215 & 0.336 &  0.393  &  \textbf{0.424} &  0.441 \\
                                &                         & - & $ 100$ & $   2$  & - & & 0.222 & 0.342 & 0.393   &  0.422 &     \textbf{0.443} \\
                                \cmidrule(r){2-12}
                                & \multirow{5}{*}{\simYP} & - & $   1$ & $   1$  & - & & \ul{\textbf{0.262}} & 0.292 &  0.305  &  0.310 &  0.320 \\
                                &                         & - & $   1$ & $  10$  & - & & 0.254 & \textbf{0.329} & 0.350   &  0.368 &  0.378 \\
                                &                         & - & $   2$ & $   5$  & - & & 0.237 & 0.312 & \textbf{0.357}   &  0.372 &  0.381 \\
                                &                         & - & $   3$ & $  20$  & - & & 0.230 & 0.312 & 0.355   &  \textbf{0.381} &  0.393 \\
                                &                         & - & $  10$ & $   1$  & - & & 0.211 & 0.273 & 0.336   &  0.374 &  \textbf{0.398} \\

        \midrule

        \multirow{4}{*}{\TptCF} & - & - &$ 160$ & - & $ 0.1$ & & \textbf{0.213} & 0.279 & 0.303 & 0.322 & 0.331\\
                                & - & - &$ 480$ & - & $ 0.9$ & & 0.189 & \textbf{0.290} & 0.320 & 0.340 & 0.355\\
                                & - & - &$ 480$ & - & $ 0.1$ & & 0.200 & 0.284 & \textbf{0.329} & 0.355 & 0.378\\
                                & - & - &$ 500$ & - & $ 0.1$ & & 0.200 & 0.282 & 0.327 & \textbf{0.357} & \textbf{0.379}\\

        \midrule
        \multirow{8}{*}{\method-\ypCF}  & \multirow{3}{*}{\simPY} & $ 0.2$  & $   1$ & $   1$ & - & & \textbf{0.247} & 0.357 & \ul{\textbf{0.426}}  &  \ul{\textbf{0.441}} & 0.464 \\
                                        &                         & $ 0.5$  & $   1$ & $   1$ & - & & 0.245 & \ul{\textbf{0.363}} & 0.422  &  0.439 & 0.464 \\
                                        &                         & $ 0.2$  & $ 100$ & $   2$ & - & & 0.226 & 0.351 & 0.404  &  0.430 & \ul{\textbf{0.467}} \\
                                        \cmidrule(r){2-12}
                                        & \multirow{5}{*}{\simYP} & $ 0.5$  & $   3$ & $   5$ & - & & \textbf{0.254} & 0.329 &  0.353  &  0.379 & 0.426 \\
                                        &                         & $ 0.1$  & $   3$ & $   2$ & - & & 0.230 & \textbf{0.346} & 0.366 &  0.402 & 0.432 \\
                                        &                         & $ 0.1$  & $   1$ & $  20$ & - & & 0.230 & 0.331 & \textbf{0.391} &  0.424 & 0.447 \\
                                        &                         & $ 0.1$  & $   1$ & $   1$ & - & & 0.222 & 0.331 & 0.383 &  \textbf{0.430} & 0.447 \\
                                        &                         & $ 0.2$  & $   1$ & $   1$ & - & & 0.222 & 0.323 & 0.378 &  0.426 & \textbf{0.449} \\

        \midrule
        \multirow{5}{*}{\method-\TptCF}  & - & $0.8$  &$  60$ & - & $ 0.4$ & & \textbf{0.228} & 0.307 & 0.335 & 0.359 & 0.379\\
                                         & - & $0.7$  &$  40$ & - & $ 0.1$ & & 0.213 & \textbf{0.312} & 0.348 & 0.376 & 0.398\\
                                         & - & $0.8$  &$ 200$ & - & $ 0.1$ & & 0.213 & 0.303 & \textbf{0.353} & 0.376 & 0.400\\
                                         & - & $0.6$  &$   5$ & - & $ 0.1$ & & 0.209 & 0.297 & 0.344 & \textbf{0.383} & 0.406\\
                                         & - & $0.1$  &$   1$ & - & $ 0.1$ & & 0.200 & 0.310 & 0.346 & 0.381 & \textbf{0.413}\\

        \bottomrule
      \end{tabular}
      \begin{tablenotes}
       \begin{scriptsize}
          \setlength\labelsep{0pt}
        \item In this table, the column ``sim'' corresponds to similarity identification
          methods; $\alpha$ is the weight on CF component in \method;
          $|\Sp|$ is the number of similar patients; $|\Sy|$ is the number of similar
          physicians; $\beta$ is the similarity threshold to identify similar terms.
          The best performance of each method under each metric is \textbf{bold}.
          The best overall performance of all methods under each metric is
          \underline{\textbf{underlined}}. 
          \par
        \end{scriptsize}
      \end{tablenotes}
  \end{threeparttable}
  \end{small}

 \vspace{-10pt}
\end{adjustwidth}
\end{table}

We compare \foMC, \ypCF, \TptCF and \method, as well as their variations,
in our experiments. 
Table~\ref{tbl:results:cuttof:0815} presents the best performance of each method. 
%
%
Overall, \method-\ypCF with \simPY is the best method because 4 out of 5 results of
\method-\ypCF with \simPY are the best among all the methods.
%
%
With parameters $\alpha$=0.2, $|\Sp|$=1 (i.e., 1 similar patient) and $|\Sy|$=1 (i.e., 1 similar physician),
\method-\ypCF with \simPY outperforms the simple \foMC at 22.3\%, 20.2\%, 26.0\%, 16.7\% and
18.1\% on HR@1, HR@2, HR@3, HR@4 and HR@5, respectively.
%
%
The second best method is \ypCF with \simPY because it has better results overall than the rest methods. With parameters $|\Sp|$=1 and
$|\Sy|$=1, \ypCF with \simPY outperforms the simple \foMC at
23.3\%, 19.5\%, 20.1\%, 10.3\% and 8.9\%
on HR@1, HR@2, HR@3, HR@4 and HR@5, respectively.
It is notable that although \ypCF is significantly better than \foMC, the best
\method-\ypCF with \simPY has a weight $\alpha$=0.2 on the \ypCF scoring component,
but a weight 1-$\alpha$=0.8 on the \foMC scoring component.
This indicates the importance of search dynamics in recommending
the next search terms.
It is also notable that the optimal \method-\ypCF with \simPY corresponds to
a very small number of similar patients ($\Sp$=1) and physicians ($\Sy$=1). This
demonstrates the effectiveness of \method-\ypCF in identifying most relevant
information and leveraging such information for term recommendation. 

%
%
The \method-\TptCF method is also slightly better than \foMC.
With parameters $\alpha$=0.1, $|\Sp|$=1 and $\beta$=0.1, \method-\TptCF outperforms
\foMC at -1.0\%, 4.4\%, 2.4\%, 0.8\% and 5.1\%
on HR@1, HR@2, HR@3, HR@4 and HR@5, respectively.
%
%
However, \method-\TptCF is significantly worse than \method-\ypCF with \simPY. The difference
between \method-\TptCF and \method-\ypCF is that in \method-\ypCF, the similarity-based
scoring component (i.e., \ypCF) does not consider search dynamics and only looks at
the search terms that have ever been searched by similar physicians on similar
patients, regardless of how such search terms transit to the search term of
interest, while \TptCF considers such transitions.
The performance difference between \method-\TptCF and \method-\ypCF may indicate that
the transition information captured in \TptCF might overlap with that captured in \foMC and
thus combining them together will not lead to substantial gains.
On the other hand, the information captured by \ypCF methods could be complementary to
that in \foMC and thus integration of \ypCF and \foMC results in significant performance
improvement.

In \method-\ypCF, \simPY is slightly better than \simYP.
%
The \simPY method first identifies patients similar to the target patient, and based on
the identified similar patients identifies physicians similar to the target physician.
The \simYP method identifies similar patients and similar physicians in the reversed order
as in \simPY. 
%
%
The better performance of \simPY over \simYP in \method-\ypCF demonstrates
that when physician search dynamics has been considered via \MC,
similar patients should be identified first and then based on identified similar patients,
similar physicians should be identified.
This may be because that when \MC already considers all patients and all
physicians (Equation~\ref{eqn:mc_p}), a more focused and more homogeneous group of
patients similar to the target patient is more critical
in order to complement to the \MC information.
Since physicians may see many patients with different diseases,
high physician similarity may be due to patients who are
different from the target patient. 
If such physicians are first selected (e.g., in \simYP),
similar patients identified from these physicians
might be very different from the target patient. 
%
%
However, when no information about all the patients and all the physicians is considered
like in \ypCF, a diverse set of physicians and patients might be beneficial, and that
could explain why in \ypCF, \simYP actually outperforms \simPY slightly.

%
Comparing \ypCF and \TptCF, it is notable that \ypCF is significantly better than
\TptCF, even though in \TptCF more patients similar to the target patient are
used to achieve its optimal performance. 
In \TptCF, only terms from similar physicians and patients
that are similar to the term of interest are considered in calculating
the scores (Equation~\ref{eqn:tdscf}). However, in \ypCF, all the terms from similar
physicians and patients are used.
The improved performance of \ypCF compared to
that of \TptCF may indicate that using more possible
terms could benefit recommendation. 
On the other hand, 
both \foMC and \TptCF consider term transitions, while \TptCF considers term transitions
only among similar terms on similar patients.
The experimental results show that \TptCF performs worse than \foMC. This may indicate that
if term transition is a major factor in determining next search term, transitions from more
diverse patients should be integrated.

%
\begin{figure}
  \centering
  \begin{minipage}{.48\textwidth}
    \centering
    \input{performance/paramter_study/all/alpha_top1.tex}
    \captionof{figure}{HR@1 over $\alpha$ values}
    \label{fig:top1}
  \end{minipage}%
\hfill
  \begin{minipage}{.48\textwidth}
    \centering
    \input{performance/paramter_study/all/alpha_top2.tex}
    \vspace{-10pt}\captionof{figure}{HR@2 over $\alpha$ values}
    \label{fig:top2}
  \end{minipage}%
  %
\end{figure}

\begin{figure}
  \centering
  \begin{minipage}{.48\textwidth}
    \centering
    \input{performance/paramter_study/all/alpha_top3.tex}
    \captionof{figure}{HR@3 over $\alpha$ values}
    \label{fig:top3}
  \end{minipage}%
  %
  \hfill
  \begin{minipage}{.48\textwidth}
    \centering
    \input{performance/paramter_study/all/alpha_top4.tex}
    \captionof{figure}{HR@4 over $\alpha$ values}
    \label{fig:top4}
  \end{minipage}%
\end{figure}
\begin{figure}
  \begin{minipage}{.5\textwidth}
    \centering
    \input{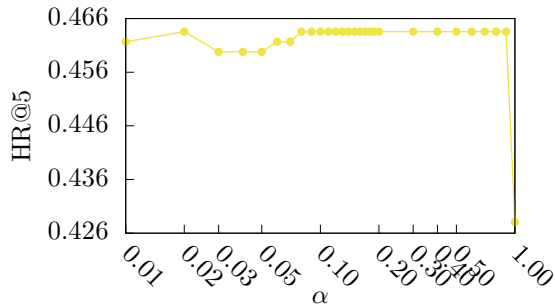}
    \captionof{figure}{HR@5 over $\alpha$ values}
    \label{fig:top5}
  \end{minipage}%
\end{figure}

Figure~\ref{fig:top1}, \ref{fig:top2}, \ref{fig:top3}, \ref{fig:top4} and \ref{fig:top5} 
present HR@1, HR@2, HR@3, HR@4 and HR@5 of \method-\ypCF with \simPY over different $\alpha$
values (Equation~\ref{eqn:score}) when $|\Sy|$ = 1 and $|\Sp|$ = 1, respectively. 
%
As the weight $\alpha$ increases from 0, that is, as the \CF takes place in
the term scoring (Equation~\ref{eqn:score}), the performance of \method in terms
of HR@1 and HR@2 generally increases.
This demonstrates the effect from \CF scoring component in \method. 
As $\alpha$ further increases, the performance in general
first gets better and then worse (except that the HR@1 performance reaches its best
at $\alpha$=1).
This indicates that the dynamic scoring component and \CF scoring component
in \method play complementary roles for recommending terms, and thus considering
their combination enables better recommendation performance than each of the two methods
alone.

\subsection*{Overall Performance on Other Cut-off Times}
\label{sec:experiments:othercutoff}

\begin{table}[!ht]
\begin{adjustwidth}{-0.4in}{0in} 
  \centering
  \caption{Statistics of INPC Dataset}
  \label{tbl:dataset_all}
  \begin{threeparttable}
    \begin{tabular}{
	@{\hspace{2pt}}l@{\hspace{2pt}}
	@{\hspace{2pt}}r@{\hspace{5pt}}
        %
        %
        @{\hspace{5pt}}c@{\hspace{5pt}}
        @{\hspace{2pt}}r@{\hspace{2pt}}
        @{\hspace{2pt}}r@{\hspace{2pt}}
        @{\hspace{5pt}}c@{\hspace{5pt}}
	@{\hspace{2pt}}r@{\hspace{2pt}}
        @{\hspace{2pt}}r@{\hspace{2pt}}
        @{\hspace{5pt}}c@{\hspace{5pt}}
	@{\hspace{2pt}}r@{\hspace{2pt}}
        @{\hspace{2pt}}r@{\hspace{2pt}}
	@{\hspace{5pt}}c@{\hspace{5pt}}
        @{\hspace{2pt}}r@{\hspace{2pt}}
        @{\hspace{2pt}}r@{\hspace{2pt}}
      }
      \toprule
      \multirow{3}{*}{statistics} & \multirow{3}{*}{INPC}
      && \multicolumn{2}{c}{\cutoff}
      && \multicolumn{2}{c}{\cutoff}
      && \multicolumn{2}{c}{\cutoff}
      && \multicolumn{2}{c}{\cutoff}
      \\
      & 
      && \multicolumn{2}{c}{(06/26/2013)}
      && \multicolumn{2}{c}{(07/18/2013)}
      && \multicolumn{2}{c}{(08/15/2013)}
      && \multicolumn{2}{c}{(09/03/2013)}\\

      \cline{4-5} \cline{7-8}\cline{10-11}\cline{13-14}
      &		
      && train & test
      && train & test
      && train & test
      && train & test\\  
      \midrule
      \#\patient 		& 13,819 	
      &&  6,669&    587
      &&  8,471&    624
      && 10,852&    472
      && 12,014&    372\\
      \#\physician 		& 2,121 	
      &&  1,267&    126
      &&  1,542&    147
      &&  1,818&    126
      &&  1,948&    105\\
      \#\term 			& 9,781 	
      &&  5,334&    665
      &&  6,550&    654
      &&  7,952&    532
      &&  8,657&    461\\
      \#\seq 		& 24,183	
      && 10,385&    648
      && 13,677&    692
      && 18,166&    535
      && 20,492&    414\\
      len(\seq) 		& 69,770	
      && 28,789&  2,568
      && 38,553&  2,506
      && 51,272&  1,831
      && 58,146&  1,482\\
      len(\seq)/\#\patient    & 5.049         
      &&  4.317&  4.375
      &&  4.551&  4.016
      &&  4.725&  3.879
      &&  4.840&  3.984\\
      len(\seq)/\#\seq	& 2.885		
      &&  2.772&  3.963
      &&  2.819&  3.621
      &&  2.822&  3.422
      &&  2.837&  3.580\\
      \bottomrule
    \end{tabular}
    \begin{tablenotes}
      \setlength\labelsep{0pt}
      \begin{footnotesize}
      \item
        In this table, \#\patient is the number of patients;
        \#\physician is the number of physicians;
        \#\term is the number of terms;
        \#\seq is the number of sequences;
        len(\seq) is total length of sequences;
        len(\seq)/\#\patient is average length of sequences per patient
        and len(\seq)/\#\seq is average length of sequences. 
        \par
      \end{footnotesize}
    \end{tablenotes}
  \end{threeparttable}

 \vspace{-10pt}
\end{adjustwidth}
\end{table}

Table~\ref{tbl:dataset_all} shows the dataset with different cut-off times
06/26/2013, 07/18/2013, 08/15/2013 and 09/03/2013. 
Table~\ref{tbl:results:cutoff:0626}, Table~\ref{tbl:results:cuttof:0718}
and Table~\ref{tbl:results:cutoff:0903} present the best performance of
all the methods for cut-off time 06/26/2013, 07/18/2013 and 09/03/2013, respectively. 
Overall, \method-\ypCF achieves the best performance over the other methods on the
different cut-off times. The trends among different methods as identified from
cut-off time 08/15/2013 remain very similar for the other cut-off times.
Note that as using later cut-off times, training data become more as shown in
Table~\ref{tbl:dataset_all}, and the performance of each method over different cut-off times
tends to become worse. For example, the performance of \foMC model decreases in general over
different cut-off times.
This may be due to the increasing heterogeneity among patients as more patients in
the system.

\begin{table}[!ht]
\begin{adjustwidth}{-0.8in}{0in} 
  \centering
  \caption{\mbox{Overall Performance Comparison with \cutoff 06/26/2013}}
  {
  \label{tbl:results:cutoff:0626}
  \begin{small}
  \begin{threeparttable}
      \begin{tabular}{
	@{\hspace{7pt}}c@{\hspace{7pt}}
        @{\hspace{7pt}}c@{\hspace{7pt}}
        @{\hspace{7pt}}r@{\hspace{7pt}}
        @{\hspace{7pt}}r@{\hspace{7pt}}
        @{\hspace{7pt}}r@{\hspace{7pt}}
        @{\hspace{7pt}}r@{\hspace{7pt}}
        @{\hspace{0pt}}r@{\hspace{0pt}}
        @{\hspace{7pt}}r@{\hspace{7pt}}
        @{\hspace{7pt}}r@{\hspace{7pt}}
        @{\hspace{7pt}}r@{\hspace{7pt}}
        @{\hspace{7pt}}r@{\hspace{7pt}}
        @{\hspace{7pt}}r@{\hspace{7pt}}
        }        
        \toprule
        method   & sim & $\alpha$ & $|\Sp|$ & $|\Sy|$ & $\beta$ & & HR@1 & HR@2 & HR@3 & HR@4 & HR@5 \\
        \midrule
        \foMC  & -                       & -          & - & - & - & & \textbf{0.205}         & \textbf{0.313}           & \textbf{0.341}           & \textbf{0.369}   & \textbf{0.381}\\

        \midrule
        \multirow{7}{*}{\ypCF}  & \multirow{3}{*}{\simPY} & - & $   4$ & $  1$& - &  &  \textbf{0.261} & 0.366 & 0.380  & 0.383  & 0.383  \\
                                &                         & - & $  50$ & $  1$& - &  &  0.259 & \textbf{0.377} & 0.398  & 0.414  & 0.418  \\
                                &                         & - & $ 100$ & $  1$& - &  &  0.250 & 0.373 & \textbf{0.403}  & \textbf{0.418}  & \textbf{0.431}  \\
                                \cmidrule(r){2-12}
                                & \multirow{4}{*}{\simYP} & - & $   2$ & $  3$& - &  &  \ul{\textbf{0.302}} & 0.350 & 0.364  & 0.369  & 0.372  \\
                                &                         & - & $   3$ & $  1$& - &  &  0.287 & \textbf{0.370} & 0.397  & 0.414  & 0.421  \\
                                &                         & - & $   5$ & $  1$& - &  &  0.279 & 0.360 & \textbf{0.401}  & \textbf{0.423}  & 0.437  \\
                                &                         & - & $  10$ & $  1$& - &  &  0.262 & 0.349 & 0.397  & 0.421  & \textbf{0.444}  \\

        \midrule

        \multirow{4}{*}{\TptCF} & - & - &$ 200$ & - & $ 0.1$ &  & \textbf{0.207} & 0.312 & 0.335 & 0.347 & 0.349\\
                                & - & - &$ 220$ & - & $ 0.1$ &  & 0.204 & \textbf{0.313} & 0.343 & 0.350 & 0.353\\
                                & - & - &$ 320$ & - & $ 0.1$ &  & 0.199 & 0.313 & \textbf{0.347} & \textbf{0.361} & 0.370\\
                                & - & - &$ 380$ & - & $ 0.1$ &  & 0.194 & 0.312 & 0.346 & 0.356 & \textbf{0.372}\\

        \midrule
        \multirow{8}{*}{\method-\ypCF}  & \multirow{3}{*}{\simPY} & $ 0.3$ & $  4$ &$ 1$&- &  & \textbf{0.262} & \ul{\textbf{0.387}} & 0.415  & 0.437  & 0.449  \\
                                        &                         & $ 0.1$ & $ 20$ &$ 1$&- &  & 0.253 & 0.377 & \ul{\textbf{0.420}}  & \textbf{0.449}  & 0.458  \\
                                        &                         & $ 0.2$ & $ 20$ &$ 1$&- &  & 0.258 & 0.381 & 0.420 & 0.449  & \textbf{0.460}  \\
                                        \cmidrule(r){2-12}
                                        & \multirow{5}{*}{\simYP} & $ 0.6$ & $  3$ & $ 10$ & - &  & \textbf{0.262} & 0.370 & 0.407  & 0.438  & 0.455  \\
                                        &                         & $ 0.4$ & $  3$ & $  1$ & - &  & 0.219 & \textbf{0.380} & 0.409  & 0.440  & 0.469  \\
                                        &                         & $ 0.2$ & $  3$ & $  4$ & - &  & 0.227 & 0.375 & \textbf{0.417}  & 0.441  & 0.463  \\
                                        &                         & $ 0.2$ & $  2$ & $  3$ & - &  & 0.216 & 0.363 & 0.412  & \ul{\textbf{0.451}}  & 0.463  \\
                                        &                         & $ 0.1$ & $  5$ & $  1$ & - &  & 0.228 & 0.373 & 0.417  & 0.443  & \ul{\textbf{0.475}}  \\

        \midrule
        \multirow{5}{*}{\method-\TptCF}  & - & $ 0.7$  &$   5$ & - & $ 0.1$ &  & \textbf{0.215} & 0.310 & 0.352 & 0.381 & 0.392\\
                                         & - & $ 0.9$  &$ 220$ & - & $ 0.1$ &  & 0.207 & \textbf{0.324} & 0.356 & 0.373 & 0.383\\
                                         & - & $ 0.8$  &$  10$ & - & $ 0.1$ &  & 0.208 & 0.312 & \textbf{0.360} & 0.384 & 0.394\\
                                         & - & $ 0.6$  &$  10$ & - & $ 0.1$ &  & 0.211 & 0.321 & 0.355 & \textbf{0.386} & 0.395\\
                                         & - & $ 0.5$  &$  10$ & - & $ 0.1$ &  & 0.208 & 0.318 & 0.353 & 0.381 & \textbf{0.397}\\

        \bottomrule
      \end{tabular}
      \begin{tablenotes}
       \begin{scriptsize}
          \setlength\labelsep{0pt}
        \item In this table, the column ``sim'' corresponds to similarity identification
          methods; $\alpha$ is the weight on CF component in \method;
          $|\Sp|$ is the number of similar patients; $|\Sy|$ is the number of similar
          physicians; $\beta$ is the similarity threshold to identify similar terms.
          The best performance of each method under each metric is \textbf{bold}.
          The best overall performance of all methods under each metric is
          \underline{\textbf{underlined}}. 
          \par
        \end{scriptsize}
      \end{tablenotes}
  \end{threeparttable}
  \end{small}
  } 

 \vspace{-10pt}
\end{adjustwidth}
\end{table}
%
\begin{table}[!ht]
\begin{adjustwidth}{-0.8in}{0in} 
  \centering
 \caption{\mbox{Overall Performance Comparison with \cutoff 07/18/2013}}
  {
  \label{tbl:results:cuttof:0718}
  \begin{small}
  \begin{threeparttable}
      \begin{tabular}{
	@{\hspace{7pt}}c@{\hspace{7pt}}
        @{\hspace{7pt}}c@{\hspace{7pt}}
        @{\hspace{7pt}}r@{\hspace{7pt}}
        @{\hspace{7pt}}r@{\hspace{7pt}}
        @{\hspace{7pt}}r@{\hspace{7pt}}
        @{\hspace{7pt}}r@{\hspace{7pt}}
        @{\hspace{0pt}}r@{\hspace{0pt}}
        @{\hspace{7pt}}r@{\hspace{7pt}}
        @{\hspace{7pt}}r@{\hspace{7pt}}
        @{\hspace{7pt}}r@{\hspace{7pt}}
        @{\hspace{7pt}}r@{\hspace{7pt}}
        @{\hspace{7pt}}r@{\hspace{7pt}}
        }        
        \toprule
        method   & sim & $\alpha$ & $|\Sp|$ & $|\Sy|$ & $\beta$ & & HR@1 & HR@2 & HR@3 & HR@4 & HR@5 \\
        \midrule
        \foMC  & -                       & -          & - & - & - &  & \textbf{0.210}         & \textbf{0.292}           & \textbf{0.325}           & \textbf{0.341}   & \textbf{0.348}\\

        \midrule
        \multirow{7}{*}{\ypCF}  & \multirow{4}{*}{\simPY} & - & $   5$ & $ 1$&- & & \textbf{0.267} & 0.347 & 0.358  & 0.364  & 0.366  \\
                                &                         & - & $  50$ & $ 1$&- & & 0.262 & \textbf{0.358} & 0.379  & 0.395  & 0.400  \\
                                &                         & - & $ 100$ & $ 1$&- & & 0.257 & 0.358 & \textbf{0.384}  & \textbf{0.402}  & 0.412  \\
                                &                         & - & $ 100$ & $ 2$&- & & 0.237 & 0.342 & 0.380  & 0.396  & \textbf{0.413}  \\
                                \cmidrule(r){2-12}
                                & \multirow{3}{*}{\simYP} & -& $   2$ & $   3$ &-& & \ul{\textbf{0.289}} & 0.337 & 0.353  & 0.357  & 0.358  \\
                                &                         & -& $   1$ & $ 100$ &-& & 0.283 & \textbf{0.345} & 0.353  & 0.357  & 0.358  \\
                                &                         & -& $  10$ & $   1$ &-& & 0.240 & 0.325 & \textbf{0.379}  & \textbf{0.410}  & \textbf{0.426}  \\

        \midrule

        \multirow{4}{*}{\TptCF} & - & - &$ 260$ & - & $ 0.1$ & & \textbf{0.210} & 0.286 & 0.301 & 0.312 & 0.329\\
                                & - & - &$ 300$ & - & $ 0.1$ & & 0.207 & \textbf{0.289} & 0.305 & 0.318 & 0.329\\
                                & - & - &$ 380$ & - & $ 0.1$ & & 0.208 & 0.288 & \textbf{0.309} & 0.324 & \textbf{0.341}\\
                                & - & - &$ 420$ & - & $ 0.1$ & & 0.208 & 0.288 & 0.308 & \textbf{0.325} & 0.340\\

        \midrule
        \multirow{8}{*}{\method-\ypCF}  & \multirow{3}{*}{\simPY} & $ 0.2$  & $   5$ & $ 1$ & -& & \textbf{0.267} & \ul{\textbf{0.364}} & 0.393  & 0.403  & 0.426  \\
                                        &                         & $ 0.1$  & $  50$ & $ 1$ & -& & 0.256 & 0.355 & \textbf{0.396}  & \textbf{0.415}  & 0.428  \\
                                        &                         & $ 0.2$  & $ 100$ & $ 1$ & -& & 0.253 & 0.360 & 0.396  & 0.413  & \textbf{0.431}  \\
                                        \cmidrule(r){2-12}
                                        & \multirow{5}{*}{\simYP} & $ 0.5$  & $   2$ & $ 3$ & -& & \textbf{0.251} & 0.347 & 0.387  & 0.408  & 0.426  \\
                                        &                         & $ 0.4$  & $   2$ & $ 4$ & -& & 0.250 & \textbf{0.351} & 0.392  & 0.413  & 0.431  \\
                                        &                         & $ 0.5$  & $   5$ & $ 4$ & -& & 0.228 & 0.341 & \ul{\textbf{0.397}}  & 0.419  & 0.441  \\
                                        &                         & $ 0.2$  & $   5$ & $ 1$ & -& & 0.228 & 0.335 & 0.389  & \ul{\textbf{0.423}}  & 0.436  \\
                                        &                         & $ 0.5$  & $  10$ & $ 4$ & -& & 0.212 & 0.315 & 0.384  & 0.412  & \ul{\textbf{0.447}}  \\

        \midrule
        \multirow{5}{*}{\method-\TptCF}  & - & $ 0.8$  &$   5$ &- & $ 0.1$ & & \textbf{0.218} & 0.292 & 0.332 & 0.351 & 0.367\\
                                         & - & $ 0.8$  &$ 300$ &- & $ 0.1$ & & 0.215 & \textbf{0.305} & 0.328 & 0.345 & 0.351\\
                                         & - & $ 0.6$  &$   5$ &- & $ 0.1$ & & 0.217 & 0.302 & \textbf{0.340} & 0.355 & 0.364\\
                                         & - & $ 0.5$  &$   5$ &- & $ 0.1$ & & 0.215 & 0.302 & 0.338 & \textbf{0.357} & 0.364\\
                                         & - & $ 0.3$  &$   1$ &- & $ 0.1$ & & 0.208 & 0.292 & 0.331 & 0.354 & \textbf{0.367}\\

        \bottomrule
      \end{tabular}
  \begin{tablenotes}
           \begin{scriptsize}
          \setlength\labelsep{0pt}
        \item In this table, the column ``sim'' corresponds to similarity identification
          methods; $\alpha$ is the weight on CF component in \method;
          $|\Sp|$ is the number of similar patients; $|\Sy|$ is the number of similar
          physicians; $\beta$ is the similarity threshold to identify similar terms.
          The best performance of each method under each metric is \textbf{bold}.
          The best overall performance of all methods under each metric is
          \underline{\textbf{underlined}}. 
          \par
        \end{scriptsize}
      \end{tablenotes}
  \end{threeparttable}
  \end{small}
  } 

 \vspace{-10pt}
\end{adjustwidth}
\end{table}
%
%
\begin{table}[!ht]
\begin{adjustwidth}{-0.8in}{0in} 
  \centering
  \caption{\mbox{Overall Performance Comparison with \cutoff 09/03/2013}}
  {
  \label{tbl:results:cutoff:0903}
  \begin{small}
  \begin{threeparttable}
      \begin{tabular}{
	@{\hspace{7pt}}c@{\hspace{7pt}}
        @{\hspace{7pt}}c@{\hspace{7pt}}
        @{\hspace{7pt}}r@{\hspace{7pt}}
        @{\hspace{7pt}}r@{\hspace{7pt}}
        @{\hspace{7pt}}r@{\hspace{7pt}}
        @{\hspace{7pt}}r@{\hspace{7pt}}
        @{\hspace{0pt}}r@{\hspace{0pt}}
        @{\hspace{7pt}}r@{\hspace{7pt}}
        @{\hspace{7pt}}r@{\hspace{7pt}}
        @{\hspace{7pt}}r@{\hspace{7pt}}
        @{\hspace{7pt}}r@{\hspace{7pt}}
        @{\hspace{7pt}}r@{\hspace{7pt}}
        }        
        \toprule
        method   & sim & $\alpha$ & $|\Sp|$ & $|\Sy|$ & $\beta$ & & HR@1 & HR@2 & HR@3 & HR@4 & HR@5 \\
        \midrule
        \foMC  & -                       & -          & - & - & - & & \textbf{0.193}         & \textbf{0.271}           & \textbf{0.304}           & \textbf{0.331}   & \textbf{0.365}\\

        \midrule
        \multirow{8}{*}{\ypCF}  & \multirow{3}{*}{\simPY} & - & $  10$ & $  1$  & - & & \textbf{0.261} & 0.326 &   0.345  &  0.355 &  0.355 \\
                                &                         & - & $  20$ & $  1$  & - & & 0.261 & \textbf{0.329} &  0.353  &  0.365 &  0.367 \\
                                &                         & - & $ 100$ & $  1$  & - & & 0.246 & 0.324 &  \textbf{0.374}  &  \textbf{0.399} &  \textbf{0.406} \\
                                \cmidrule(r){2-12}
                                & \multirow{5}{*}{\simYP} & - & $   1$ & $  1$  & - & & \ul{\textbf{0.278}} & 0.329 &  0.350  &  0.365 &  0.365 \\
                                &                         & - & $   2$ & $  3$  & - & & 0.271 & \textbf{0.336} & 0.360   &  0.379 &  0.384 \\
                                &                         & - & $  10$ & $  1$  & - & & 0.234 & 0.304 & \textbf{0.372}   &  0.391 &  0.406 \\
                                &                         & - & $   5$ & $  1$  & - & & 0.242 & 0.331 & 0.362  & \textbf{0.396} &  0.408 \\
                                &                         & - & $  10$ & $ 20$  & - & & 0.222 & 0.300 & 0.360  & 0.389 &  \textbf{0.413} \\

        \midrule

        \multirow{3}{*}{\TptCF} & - & - &$ 180$ & - & $ 0.1$ & & \textbf{0.184} & 0.246 & 0.271 & 0.290 & 0.304\\
                                & - & - &$ 320$ & - & $ 0.1$ & & 0.179 & \textbf{0.266} & 0.295 & 0.309 & 0.326\\
                                & - & - &$ 500$ & - & $ 0.1$ & & 0.174 & 0.261 & \textbf{0.312} & \textbf{0.338} & \textbf{0.353}\\

        \midrule
        \multirow{9}{*}{\method-\ypCF}  & \multirow{4}{*}{\simPY} & $0.2$  & $  10$ & $ 1$ & - & & \textbf{0.263} & 0.336 & 0.377  &  0.389 & 0.411 \\
                                        &                         & $0.1$  & $  10$ & $ 1$ & - & & 0.261 & \textbf{0.338} & 0.377  &  0.389 & 0.411 \\
                                        &                         & $0.1$  & $ 100$ & $ 1$ & - & & 0.234 & 0.331 & \textbf{0.382}  &  \ul{\textbf{0.411}} & 0.425 \\
                                        &                         & $0.2$  & $ 100$ & $ 1$ & - & & 0.246 & 0.331 & 0.382  & 0.408  & \textbf{0.428} \\
                                        \cmidrule(r){2-12}
                                        & \multirow{5}{*}{\simYP} & $0.4$  & $   3$ & $ 2$ & - & & \textbf{0.242} & 0.319 &  0.355  &  0.386 &  0.423 \\
                                        &                         & $0.4$  & $   2$ & $ 1$ & - & & 0.234 & \ul{\textbf{0.343}} & 0.384 &  0.391 & 0.418 \\
                                        &                         & $0.3$  & $   3$ & $ 2$ & - & & 0.234 & 0.336 & \ul{\textbf{0.389}} &  0.396 & 0.423 \\
                                        &                         & $0.2$  & $   4$ & $ 5$ & - & & 0.220 & 0.333 & 0.374 &  \textbf{0.403} & 0.425 \\
                                        &                         & $0.1$  & $   2$ & $ 2$ & - & & 0.208 & 0.312 & 0.362 &  0.391 & \ul{\textbf{0.435}} \\

        \midrule
        \multirow{5}{*}{\method-\TptCF}  & - & $0.8$  &$  40$ & - & $ 0.1$ & & \textbf{0.208} & 0.292 & 0.326 & 0.348 & 0.374\\
                                         & - & $0.8$  &$  20$ & - & $ 0.1$ & & 0.198 & \textbf{0.292} & 0.321 & 0.345 & 0.379\\
                                         & - & $0.9$  &$ 460$ & - & $ 0.1$ & & 0.181 & 0.271 & \textbf{0.338} & 0.365 & 0.382\\
                                         & - & $0.9$  &$ 480$ & - & $ 0.1$ & & 0.184 & 0.271 & 0.333 & \textbf{0.367} & 0.382\\
                                         & - & $0.1$  &$   5$ & - & $ 0.1$ & & 0.198 & 0.278 & 0.319 & 0.350 & \textbf{0.389}\\

        \bottomrule
      \end{tabular}
      \begin{tablenotes}
           \begin{scriptsize}
          \setlength\labelsep{0pt}
        \item In this table, the column ``sim'' corresponds to similarity identification
          methods; $\alpha$ is the weight on CF component in \method;
          $|\Sp|$ is the number of similar patients; $|\Sy|$ is the number of similar
          physicians; $\beta$ is the similarity threshold to identify similar terms.
          The best performance of each method under each metric is \textbf{bold}.
          The best overall performance of all methods under each metric is
          \underline{\textbf{underlined}}. 
          \par
        \end{scriptsize}
      \end{tablenotes}
  \end{threeparttable}
  \end{small}
  } 

 \vspace{-10pt}
\end{adjustwidth}
\end{table}

\subsection*{Similarity Analysis}
\label{sec:experiments:sim}

\begin{figure}[!h]
  \centering
  \begin{minipage}{.48\textwidth}
    \centering
    \input{graphs/phy_similarity_hist.tex}
    \captionof{figure}{Physician-physician similarity distribution}
    \label{fig:phy_phy_sim_graph}
  \end{minipage}
  %
%
\hfill
  \centering
  \begin{minipage}{.48\textwidth}
    \centering
    \input{graphs/pat_similarity_hist.tex}
    \captionof{figure}{Patient-patient similarity distribution}
    \label{fig:pat_pat_sim_graph}
  \end{minipage}
  %
\end{figure}
%
%

Figure~\ref{fig:phy_phy_sim_graph} and \ref{fig:pat_pat_sim_graph} 
present the distribution of non-zero physician-physician similarities (\simY) and
patient-patient similarities (\simP), respectively.
For \simY, 5.65\% of physician-physician similarities are non-zero, and
80.98\% of the non-zero similarities are less than or equal to 0.2.
For \simP, 2.65\% of the patient-patient similarities are non-zero, and
77.05\% of the non-zero similarities are less than or equal to 0.5.
Specially, there are some patients whose similarities with one another are relatively high
(i.e., the peaks in Figure~\ref{fig:pat_pat_sim_graph} on larger \simP values). 
This also explains the advantages of \simPY over \simYP and their performance in
Table~\ref{tbl:results:cuttof:0815}, because more patients with higher \simP to the target patient
provide better opportunities for \method to identify relevant information from such similar
patients.

\begin{figure}[!h]
  \centering
    \centering
   \scalebox{0.75}{\input{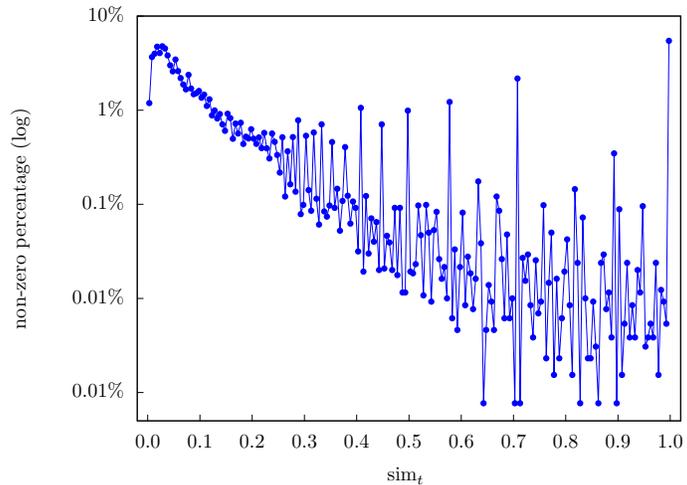}}
    \captionof{figure}{Term-term similarity distribution}
    \label{fig:term_term_sim_graph}
\end{figure}

Figure ~\ref{fig:term_term_sim_graph}
presents the distribution of non-zero term-term similarities (\simT). 
For \simT, only 0.28\% of term-term similarities are non-zero, and
78.36\% of the
non-zero similarities are less than or equal to 0.3.
%





\section*{Conclusions}
\label{sec:conclusion}

In this manuscript, we presented our new dynamic and multi-collaborative filtering
method \method to recommend search terms relevant to patients for physicians.
\method combines a dynamic first-order Markov chain model and a multi-collaborative
filtering model in order to score and prioritize search terms.
The collaborative
filtering model leverages the key idea originating from Recommender Systems
research, and uses patient similarities, physician similarities and term similarities
to score potential search terms.
The linear combination of the dynamic-based scoring
and the multi-collaborative filtering-based scoring is able to produce high quality
recommendations that are most relevant to the patients and that are most interested
to physicians.

\section*{Acknowledgment}
\label{sec:ack}


This project was made possible, in part, by support from
the National Science Foundation under Grant Number IIS-
1855501 and IIS-1827472, and from National Library of
Medicine under Grant Number 1R01LM012605-01A1. Any
opinions, findings, and conclusions or recommendations expressed
in this material are those of the authors and do not
necessarily reflect the views of the funding agencies.

\nolinenumbers

%
%
%

\bibliographystyle{plos2015}
\bibliography{paper.bib}

\begin{thebibliography}{10}

\bibitem{Ricci2015}
Ricci F, Rokach L, Shapira B.
\newblock Recommender systems handbook.
\newblock 2nd ed. Springer Publishing Company, Incorporated; 2015.

\bibitem{Ning2015}
Ning X, Desrosiers C, Karypis G.
\newblock A comprehensive survey of neighborhood-based recommendation methods.
\newblock In: Ricci F, Rokach L, Shapira B, editors. Recommender Systems
  Handbook. Boston, MA: Springer US; 2015. p. 37--76.

\bibitem{Zhang2017}
Zhang S, Yao L, Sun A. Deep learning based recommender system: a survey and new
  perspectives; 2017.

\bibitem{zhang2014latent}
Zhang C, Wang K, Yu H, Sun J, Lim EP.
\newblock Latent factor transition for dynamic collaborative filtering.
\newblock In: Proceedings of the 2014 SIAM International Conference on Data
  Mining. SIAM; 2014. p. 452--460.

\bibitem{sahoo2012hidden}
Sahoo N, Singh PV, Mukhopadhyay T.
\newblock A hidden markov model for collaborative filtering.
\newblock Mis Quarterly. 2012; p. 1329--1356.

\bibitem{sun2012dynamic}
Sun JZ, Varshney KR, Subbian K.
\newblock Dynamic matrix factorization: a state space approach.
\newblock In: Acoustics, Speech and Signal Processing (ICASSP), 2012 IEEE
  International Conference on. IEEE; 2012. p. 1897--1900.

\bibitem{sun2014collaborative}
Sun JZ, Parthasarathy D, Varshney KR.
\newblock Collaborative kalman filtering for dynamic matrix factorization.
\newblock IEEE Trans Signal Processing. 2014;62(14):3499--3509.

\bibitem{Luo2015}
Luo D, Xu H, Zhen Y, Ning X, Zha H, Yang X, et~al.
\newblock Multi-task multi-dimensional Hawkes processes for modeling event
  sequences.
\newblock In: Proceedings of the 24th International Joint Conference on
  Artificial Intelligence. IJCAI'15; 2015. p. 3685--3691.

\bibitem{xiong2010temporal}
Xiong L, Chen X, Huang TK, Schneider J, Carbonell JG.
\newblock Temporal collaborative filtering with bayesian probabilistic tensor
  factorization.
\newblock In: Proceedings of the 2010 SIAM International Conference on Data
  Mining. SIAM; 2010. p. 211--222.

\bibitem{Pfeifer2014}
Wiesner M, Pfeifer D.
\newblock Health recommender systems: concepts, requirements, technical basics
  and challenges.
\newblock International Journal of Environmental Research and Public Health.
  2014;11(3):2580--2607.
\newblock doi:{10.3390/ijerph110302580}.

\bibitem{Guo2016}
Guo L, Jin B, Yao C, Yang H, Huang D, Wang F.
\newblock Which doctor to trust: a recommender system for identifying the right
  doctors.
\newblock Journal of medical Internet research. 2016;18(7).

\bibitem{Jiang2014}
Jiang H, Xu W.
\newblock How to find your appropriate doctor: an integrated recommendation
  framework in big data context.
\newblock In: Computational Intelligence in Healthcare and e-health (CICARE),
  2014 IEEE Symposium on. IEEE; 2014. p. 154--158.

\bibitem{Wu2015}
Zhang Q, Zhang G, Lu J, Wu D.
\newblock A framework of hybrid recommender system for personalized clinical
  prescription.
\newblock In: 2015 10th International Conference on Intelligent Systems and
  Knowledge Engineering (ISKE); 2015. p. 189 -- 195.

\bibitem{Bao2016}
Bao Y, Jiang X.
\newblock An intelligent medicine recommender system framework.
\newblock In: Industrial Electronics and Applications (ICIEA), 2016 IEEE 11th
  Conference on. IEEE; 2016. p. 1383--1388.

\bibitem{Grasser2016}
Gr{\"a}{\ss}er F, Malberg H, Zaunseder S, Beckert S, K{\"u}ster D, Schmitt J,
  et~al.
\newblock Application of recommender system methods for therapy decision
  support.
\newblock In: 2016 IEEE 18th International Conference on e-Health Networking,
  Applications and Services (Healthcom); 2016. p. 1--6.

\bibitem{Duan2011}
Duan L, Street WN, Xu E.
\newblock Healthcare information systems: data mining methods in the creation
  of a clinical recommender system.
\newblock Enterprise Information Systems. 2011;5(2):169--181.

\bibitem{Moffett2011}
Moffett P, Moore G.
\newblock The standard of care: legal history and definitions: the bad and good
  news.
\newblock Western Journal of Emergency Medicine. 2011;12(1):109.

\bibitem{Lewis2007}
Lewis MH, Gohagan JK, Merenstein DJ.
\newblock The locality rule and the physician's dilemma: local medical
  practices vs the national standard of care.
\newblock JAMA. 2007;297(23):2633--2637.

\bibitem{norris1998markov}
Norris JR.
\newblock Markov chains.
\newblock 2. Cambridge university press; 1998.

\bibitem{Aggarwal2012}
Aggarwal CC, Zhai C.
\newblock Mining text data.
\newblock Springer Science \& Business Media; 2012.

\end{thebibliography}


\begin{thebibliography}{10}
\bibitem{bib1}
Conant GC, Wolfe KH.
\newblock {{T}urning a hobby into a job: how duplicated genes find new
  functions}.
\newblock Nat Rev Genet. 2008 Dec;9(12):938--950.
\bibitem{bib2}
Ohno S.
\newblock Evolution by gene duplication.
\newblock London: George Alien \& Unwin Ltd. Berlin, Heidelberg and New York:
  Springer-Verlag.; 1970.
\bibitem{bib3}
Magwire MM, Bayer F, Webster CL, Cao C, Jiggins FM.
\newblock {{S}uccessive increases in the resistance of {D}rosophila to viral
  infection through a transposon insertion followed by a {D}uplication}.
\newblock PLoS Genet. 2011 Oct;7(10):e1002337.
\end{thebibliography}

\end{document}